\documentstyle[12pt,epsf]{article}

\newcommand\be{\begin{equation}}
\newcommand\ee{\end{equation}}
\newcommand\bea{\begin{eqnarray}}
\newcommand\eea{\end{eqnarray}}
\newcommand\beas{\begin{eqnarray*}}
\newcommand\eeas{\end{eqnarray*}}

\newcommand\lae{\stackrel{<}{\sim}}
\newcommand\gae{\stackrel{>}{\sim}}

\newcommand{\NPB}[1]{{\it Nucl. Phys.}\ {\bf B{#1}}}
\newcommand{\PLB}[1]{{\it Phys. Lett.}\ {\bf B{#1}}}
\newcommand{\PRD}[1]{{\it Phys. Rev.}\ {\bf D{#1}}}
\newcommand{\PRL}[1]{{\it Phys. Rev. Lett.}\ {\bf #1}}

\def\st{\sin\theta}
\def\ct{\cos\theta}
\def\sp{\sin\phi}
\def\cp{\cos\phi}

\def\gp{g^\prime}
\def\gq{g_{light}}
\def\gl{g_{heavy}
}

\newcommand\iip{\ \ .}

\begin{document}

\begin{titlepage}
\def\thepage {}     

\title{New Gauge Interactions and Single Top Quark Production}
\author{Elizabeth H. Simmons\thanks{e-mail address:
simmons@bu.edu} \\
Department of Physics, Boston University \\
590 Commonwealth Ave., Boston  MA  02215}

\date{\today}
\maketitle

\bigskip
\begin{picture}(0,0)(0,0)
\put(295,250){BUHEP-96-37}
\put(295,235){hep-ph/9612402}
\end{picture}
\vspace{24pt}

\begin{abstract}
  Extensions of the standard model that include new $W$ bosons or
  extended technicolor gauge bosons can predict sizeable changes in the rate
  of single top quark production, even when constrained to be consistent
  with precision electroweak data.  We analyze the
  fractional change in the rate of single top quark production for several
  classes of models and determine which ones predict an effect visible in
  the Tevatron collider's Run 3.
  
\pagestyle{empty}
\end{abstract}

\end{titlepage}


\section{Introduction}
\label{sec:singletop}
\setcounter{equation}{0}

It has been suggested \cite{wilstel} that a sensitive measurement of the
$Wtb$ coupling can be made at the Tevatron collider by studying single
top production through quark/anti-quark annihilation $(q\bar q'\to W \to
t b)$ \cite{stp}, and normalizing to the Drell-Yan process $(q\bar q' \to
W q\to \ell\nu)$ to control theoretical systematic uncertainties (e.g.
in the initial parton distributions).  This method should be more
precise than alternative methods involving single top production via
$W$-gluon fusion \cite{wgf}, because there is no similar way to
eliminate the uncertainty associated with the gluon distribution
function.

In the standard model, the ratio of single top production and Drell-Yan
cross-sections
\be {\sigma(q\bar q' \to W \to t b) \over {\sigma(q\bar q'\to W \to
    \ell\nu)}} \equiv R^{SM}_\sigma 
\label{oneptone}
\ee 
is proportional to the top quark decay width $\Gamma(t\to W b)$ and,
therefore, to $\vert V_{tb}\vert^2$.  Recent work \cite{heinson} has
shown that with a 30 fb$^{-1}$ data sample from Run 3 at the Tevatron
with $\sqrt{s} = 2$ TeV it should be possible to use single top-quark
production to measure $\Delta R_\sigma / R_\sigma$, and hence $\vert
V_{tb} \vert^2$ in the standard model, to an accuracy of at least
$\pm$8\%.  By that time, the theoretical accuracy in the standard
model calculation is projected to become at least this good \cite{wilsm}

Many theories of physics beyond the standard model include new particles
or interactions that can contribute to the rates of single top-quark
production or the Drell-Yan process, thereby altering the predicted
value of $R_\sigma$.  If the resulting fractional change in the
cross-section ratio
\be
 {R_{\sigma} - R_{\sigma}^{SM} \over {R_\sigma^{SM}}} \equiv \Delta
 R_\sigma / R_\sigma
\ee
is at least 16\%, it should be detectable in Run 3.  By considering the
size of $\Delta R_\sigma / R_\sigma$ predicted by different types of new
physics, we can assess the likelihood that the
measurement of single top-quark production will help distinguish among
various classes of models.

This paper focuses largely on models that include new gauge bosons
coupled to the ordinary fermions.  The models we consider alter
$R_\sigma$ in two distinct ways, each corresponding to the presence of a
specific type of extra gauge boson.  In models of dynamical electroweak
symmetry breaking, exchange of new gauge bosons can make a large direct
correction to the $Wtb$ vertex.  In models with enlarged weak gauge
groups, two sets of $W$ bosons can be present; both sets contribute to
the cross-sections and mixing between the two sets alters the couplings
of the lighter $W$ state to fermions.  Sections 2 and 3 examine models
of dynamical electroweak symmetry breaking with (3) or without (2) extra
weak gauge bosons.  In section 4, models with light Higgs bosons and
extra weak gauge bosons are discussed.  The last section summarizes our
findings and compares the results to those obtained by others for models
of non-standard physics that do not include new gauge interactions.

\section{Ordinary Extended Technicolor}
\label{sec:comETC}
\setcounter{equation}{0}

In ordinary extended technicolor (ETC) models \cite{ETC}, the extended
technicolor gauge group commutes with the weak gauge group.  Such models
have no extra weak gauge bosons, so that the only effect on $R_\sigma$
comes from a direct ETC correction to the $Wtb$ vertex.

In order to calculate this correction, we use the methods established
for finding how ETC gauge boson exchange alters the $Zbb$ coupling 
\cite{cETC}.  Recall that the size of the effect on $Zbb$ is set by the
top quark mass. In ordinary ETC models, the top quark mass is generated
by four-fermion operators induced by the exchange of ETC gauge bosons:
\be 
{\cal L}^{ETC}_{4f} = - {2\over f^2}\left(\xi\bar\psi_L \gamma^\mu
T_L + {1\over \xi}\bar t_R \gamma^\mu U_R\right) \left(\xi \bar T_L
\gamma_\mu \psi_L + {1\over \xi}\bar U_R \gamma_\mu t_R \right) ~,
\label{fourf}
\ee
where $\xi$ is a model-dependent Clebsch; the top-bottom doublet
$\psi_L = (t,b)_L$  and the technifermion doublet $T_L = (U,D)_L$ are
weak doublets; and the scale $f$ is related (in the absence of
fine-tuning) as $f = 2 M/g$ to the ETC boson's mass and gauge
coupling.  When the technifermions
condense, the $LR$ cross-terms in the operator (\ref{fourf}) produce a
top quark mass \cite{cETC}
\be
m_t\approx {g^2 4\pi f_Q^3 \over M^2} ~,
\label{mt}
\ee
where the numerator contains an estimate of the technifermion
condensate (using dimensional analysis \cite{dimanal}) and $f_Q$ is
the Goldstone boson decay constant associated with the technifermions
which help provide a mass to the top quark.  In a one-doublet technicolor
model, $f_Q = v = 250$ GeV.  

The purely left-handed piece of operator (\ref{fourf}) affects the the
$Zbb$, $Ztt$ and $Wtb$ vertices.  As shown in ref. \cite{cETC}, that
left-handed interaction is equivalent to
\be
{\xi^2\over 2} {g^2 f_Q^2\over{M^2}} \bar \psi_L \left[ {e\over {\st\ct}}
Z\!\!\!\!/ {\tau_3\over 2} + {e\over
  {\sqrt{2}\st}}(W^+\!\!\!\!\!\!\!\!/\ \tau^+ +
W^-\!\!\!\!\!\!\!\!/\ \tau^-)\right] \psi_L .
\label{eqvto}
\ee
Hence the $Wtb$ coupling is shifted by (taking $V_{tb} = 1$)
\be
(\delta g)^{ETC} = - {\xi^2\over 2} {g^2 f_Q^2\over M^2}
{e\over {\sqrt{2}\st}}  = - {\xi^2\over 2} {m_t\over{4\pi f_Q}} 
{e\over {\sqrt{2}\st}} 
\ee
The effect of the shifted coupling on the ratio of cross-sections $R_\sigma$ is
\be
{\Delta R_\sigma\over{R_\sigma}} \approx {2\over g} \left[ (\delta g)^{ETC}
\right] \approx - 5.6\% \xi^2 \left({250{\rm GeV}\over{f_Q}}\right)
  \left({m_t\over{175 {\rm GeV}}}\right) \iip
\label{threpthre}
\ee 
Since $\xi^2$ is generally of order 1, this lies well below the
projected sensitivity of the Tevatron's Run 3.  Ordinary extended
technicolor models, then, do not predict a visible change to the rate of
single top quark production.

Note that operator \ref{eqvto} also induces a fractional shift in $R_b$
\cite{cETC}
\be
{\Delta R_b\over{R_b}} \approx {2\over g} \left[ (\delta g)^{ETC}
\right] \approx - 5.6\% \xi^2 \left({250{\rm GeV}\over{f_Q}}\right)
  \left({m_t\over{175 {\rm GeV}}}\right) \iip
\ee 
of the same size as $\Delta R_\sigma / R_\sigma$.  The current LEP value
\cite{rbtod} of $R_b$ (0.2179$\pm$0.0012) lies close enough to the
standard model prediction (0.2158) that a 5\% reduction in $R_b$ is
excluded at better than the 10$\sigma$ level.  Moreover, attempts to
increase $\Delta R_\sigma / R_\sigma$ in ordinary technicolor models may
cause the predicted value of $R_b$ to deviate still further from the
measured value.\footnote{A recent effective-Lagrangian analysis of a
non-standard contribution to the $Zbb$ and $Wtb$ vertices \cite{dazh}
similarly finds that a large shift in $R_b$ is the price of a
visible shift in $R_\sigma$.}

An interesting extension of ordinary extended technicolor models are
topcolor-assisted technicolor models \cite{TC2} in which technicolor is
responsible for most of the electroweak symmetry breaking and new strong
dynamics coupled to the top and bottom quarks generates most of the top
quark mass.  The ETC sector of such models will have an effect on
$R_\sigma$ of the form described above -- but the size of the effect is
modified by the differing values of $f_Q$ and $m_t^{ETC}$ (the part of
the top quark mass contributed by the ETC sector).  Using
typical \cite{TC2} values $f_Q \sim$ 240 GeV and $m_t^{ETC} \sim$ 1 GeV
we find that ETC-induced shift $\Delta R_\sigma/R_\sigma$ is a fraction
of a percent.  Exchange of the new `coloron' gauge bosons between the
$t$ and $b$ quarks can additionally modify the $Wtb$ vertex;
extrapolating from the results of ref. \cite{TC2hz}, which considered
similar effects on the $Zb\bar{b}$ vertex, we estimate that this
contributes at most a few percent to $R_\sigma$ at the
momentum-transfers where most of the single top production occurs.  Thus
topcolor-assisted technicolor models do not predict a visible alteration
of $R_\sigma$.

\section{Non-Commuting Extended Technicolor}
\label{sec:NonComETC}
\setcounter{equation}{0}

In ``non-commuting'' extended technicolor models, the gauge
groups for extended technicolor and for the weak interactions do not
commute.  In other words, $SU(2)_L$ is partially embedded in the ETC
gauge group and ETC gauge bosons carry weak charge.  As a result the
models include both ETC gauge bosons and an extra set of weak gauge
bosons \cite{NCETC}.

The pattern of gauge symmetry breaking required in non-commuting ETC
models generally involves three scales (rather than just two as in
ordinary ETC) to provide masses for one family of ordinary
fermions:

\vspace{1cm}
\begin{center}
$G_{ETC}  \otimes SU(2)_{light} \otimes U(1)' $
\end{center}
\vspace{-15pt}
\begin{center}
$\downarrow \ \ \ \ \ f $
\end{center}
\vspace{-15pt}
\begin{center}
$G_{TC} \otimes SU(2)_{heavy}  \otimes SU(2)_{light} \otimes U(1)_Y $
\end{center}
\vspace{-15pt}
\begin{center}
$\downarrow\ \ \ \ \ u $
\end{center}
\vspace{-15pt}
\begin{center}
$G_{TC}  \otimes SU(2)_{L} \otimes U(1)_Y$
\end{center}
\vspace{-15pt}
\begin{center}
$\downarrow\ \ \ \ \ v $
\end{center}
\vspace{-15pt}
\begin{center}
$G_{TC}  \otimes U(1)_{em}$,
\end{center}
The $SU(2)_{heavy}$ gauge group (a subgroup of $G_{ETC}$) is effectively
the weak gauge group for the third generation, while the $SU(2)_{light}$
is the weak gauge group for the two light generations.  Keeping the two
$SU(2)$ groups distinct at high energies allows a range of fermion
masses to be generated.  The two $SU(2)$'s break to a diagonal $SU(2)_L$
subgroup (which we identify with $SU(2)_{weak}$) at the scale $u$,
thereby preserving the observed low-energy universality of the weak
interactions.  The final electroweak symmetry breaking is
accomplished dynamically at the weak scale $v$.

The two simplest possibilities for the $SU(2)_{heavy} \times
 SU(2)_{light}$ transformation properties of the order parameters that
mix and break the $SU(2)$ groups are \cite{NCETC}
\be
\langle \varphi \rangle \sim (2,1)_{1/2},\ \ \ \ \langle
\sigma\rangle \sim (2,2)_0 ~,\ \ \ \ \ \ \ ``{\rm heavy\ case}"~,
\ee
and
\be
\langle \varphi \rangle \sim (1,2)_{1/2},\ \ \ \ \langle
\sigma\rangle \sim (2,2)_0 ~,\ \ \ \ \ \ \ ``{\rm light\ case}"~,
\ee
where order parameter $\langle\varphi\rangle$ breaks
$SU(2)_L$ while $\langle\sigma\rangle$ mixes
$SU(2)_{heavy}$ with $SU(2)_{light}$.  We refer to these two
possibilities as ``heavy'' and ``light'' according to whether 
$\langle\varphi\rangle$
transforms non-trivially under $SU(2)_{heavy}$ or $SU(2)_{light}$.
In the heavy case \cite{NCETC}, the technifermion condensate
responsible for providing mass for the third generation of quarks and
leptons is also responsible for the bulk of electroweak symmetry
breaking (as measured by the contribution made to the $W$ and $Z$
masses).  In the light case, the physics responsible for providing mass
for the third generation {\it does not} provide the bulk of electroweak
symmetry breaking.  

\subsection{Direct ETC effects on the $Wtb$ vertex}

{\it A priori}, it appears that the $Wtb$ vertex may be affected by both
ETC gauge boson exchange and weak gauge boson mixing.  However, a
closer look at the operator that gives rise to the top quark mass
demonstrates that there are no direct ETC contributions to the $Wtb$
vertex of order $m_t/4\pi v$ in non-commuting ETC models.  The
left-handed third generation quarks $\psi_L = (t,b)_L$ and
right-handed technifermions $T_R = (U,D)_R$ are doublets under
 $SU(2)_{heavy}$ while the left-handed technifermions are
$SU(2)_{heavy}$ singlets.  The four-fermion interaction whose
left-right interference piece gives rise to the top quark mass may be
written as \cite{NCETC}
\be
{\cal L}^{ncETC}_{4f} = - {2\over f^2}\left(\xi\bar\psi_L \gamma^\mu U_L +
{1\over \xi}\bar t_R \gamma^\mu T_R\right) \left(\xi \bar U_L \gamma_\mu
\psi_L + {1\over \xi}\bar T_R \gamma_\mu t_R \right) ~,
\label{fourfp}
\ee 
where $\xi$ is a model-dependent Clebsch.  This is the operator that
can potentially alter couplings between the weak bosons and the
third-generation quarks by an amount of order $m_t/4\pi v$.  However,
because the left-left piece of this operator includes $(t_l, b_l, U_L)$
but not $D_L$ and because its purely right-handed piece contains
$(t_R,U_R,D_R)$ but not $b_R$, this operator does {\bf not} contribute
to the $Wtb$ vertex.

This is in contrast to the result for $R_b$ where a similar
operator involving electrically neutral currents does affect the $Zb\bar
b$ coupling \cite{NCETC}.

\subsection{Extra weak gauge bosons in non-commuting ETC}
The extra set of weak gauge bosons in non-commuting ETC models affects
$R_\sigma$ both because there are now two $W$ bosons participating in
the scattering process and because gauge boson mixing alters the light
$W$ boson's couplings to fermions. We summarize here the properties of
the $W$ bosons (mass, couplings, width) that are directly relevant to
calculating $\Delta R_\sigma / R_\sigma$.  Further details are in ref. 
\cite{NCETC}.

The electromagnetic gauge group $U(1)_{em}$ is generated by $Q=
T_{3l}+T_{3h}+Y$ and the associated photon eigenstate can be written as
\be
A^\mu = \st \sp \,W_{3l}^\mu + \st \cp \,W_{3h}^\mu +\ct X^\mu~,
\label{pho}
\ee 
where $\theta$ is the weak angle and $\phi$ is an additional mixing
angle.  In terms of the electric charge and these mixing angles, the
gauge couplings of the original $SU(2)_{heavy}\times
SU(2)_{light}\times U(1)_Y$ gauge groups are
\be 
\gq = {e\over s\st}~,\ \ \ \ \gl = {e\over c\st}~,\ 
\ \ \ \gp = {e\over \ct}~, 
\ee 
where $s\equiv\sp\,$ and $c\equiv\cp$.  

It is convenient to discuss the $W$  mass eigenstates in the
rotated basis 
\be 
W^{\pm}_1 = s\,W^{\pm}_l+c\,W^{\pm}_h\ \ \ \ \ \ \ \ \ \ \ \ \ \ \ \
W^{\pm}_2 = 
c\,W^{\pm}_l-s\,W^{\pm}_h\,,
\ee 
so the gauge covariant derivatives
separate into standard and non-standard parts 
\be 
D^\mu =\partial^\mu + ig\left( T_l^\pm + T_h^\pm \right)
W^{\pm\,\mu}_1 +ig\left( {c \over s}T_l^\pm - {s \over c}T_h^\pm \right)
W^{\pm\,\mu}_2 + ...
\label{stnst}
\ee
with $g\equiv {e \over \st}$.  By diagonalizing the mass matrix of
the $W$ bosons in the limit where $u^2/v^2 \equiv x $ is large,
we can find the form of the light and heavy mass eigenstates $W^L$ and $W^H$.  For
the heavy case of non-commuting ETC, we have
\be 
W^L \approx W_1+{c s^3 \over
  {x}}\,W_2~, \ \ \ \ \ W^H\approx W_2-{c s^3 \over
  {x}}\,W_1~.
\label{dgzi}
\ee
In the light case, we have mass eigenstates
\be
W^L \approx  W_1-{c^3 s \over {x}}\,W_2~, \ \ \ \ \ 
W^H \approx W_2+{c^3 s \over {x}}\,W_1~.
\label{dgzii}
\ee
In either case, the mass of the heavy $W$ boson is approximately
given by
\be
M_{W^H} \approx
{\sqrt{x}\over s c} M_W
\label{mwzrat}
\ee
where $M_W$ is the tree-level standard model mass of the $W$ boson.
The tree-level (pole) width of the heavy $W$ boson is 
\be
\Gamma_{W^H} = {g^2\over {12\pi^2}}\left({2 c^2\over s^2} + {s^2\over
c^2}\right) M_{W^H} \iip
\label{mwgam}
\ee

\subsection{Results}

Using the information on the mass, width and couplings of the $W$ bosons
from the previous sections, we found the size of $\Delta R_\sigma/
R_\sigma$ in both the heavy and light cases of non-commuting ETC.
Details of the calculation are given in the appendix.  We used results
from ref. \cite{NCETC} to fix the 95\% c.l. experimental constraints on
the model from low-energy and LEP precision electroweak measurements;
these are stronger 
than limits from direct searches \cite{dirs} for heavy weak
bosons at FNAL.  Physically speaking, the constraints tell us the
lightest possible value of $M_{W^H}$ for any given value of
$\sin^2_\phi$, i.e. the value of $M_{W^H}$ yielding the largest $\Delta
R_\sigma/ R_\sigma$.

By checking the maximum $\Delta R_\sigma/ R_\sigma$ in the
experimentally allowed region for heavy case non-commuting ETC, we
find that $\vert\Delta R_\sigma/ R_\sigma\vert$ never exceeds 9\%.  This
means that the shift in the rate of single top quark production is never
large enough to be clearly visible at Tev33.

Repeating the exercise for the light case of non-commuting ETC leads to
a very different conclusion.  The pattern of shifts in the predicted
values of various electroweak observables has been found \cite{NCETC} to
allow the extra weak bosons in light non-commuting ETC to be as light as
400 GeV.  Since lighter extra bosons produce larger shifts in
$R_\sigma$, there is a significant overlap between the experimentally
allowed portion of parameter space and the region in which $\vert \Delta
R_\sigma/ R_\sigma\vert \geq 16$\%, as shown in Figure 1.  In fact,
the predicted fractional shift in $R_\sigma$ is greater than  24\% for much
of this overlap region.  More
precisely, the shift in $R_\sigma$ is towards values exceeding
$R_\sigma^{SM}$, so that non-commuting ETC models with the ``light''
symmetry breaking pattern predict a visible {\bf increase} in the rate
of single top-quark production.

\begin{figure}[htb]
\epsfxsize 10cm \centerline{\epsffile{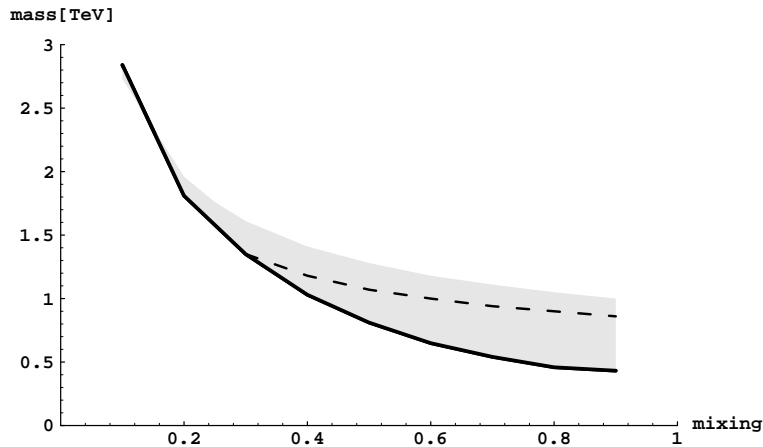}}
\caption[ncetclim]{Region (shaded) where light-case non-commuting ETC models
  predict a visible increase ($\Delta R_\sigma/ R_\sigma \geq 16$\%) in
  single top quark production at TeV33.  The dark line marks the lower
  bound (at 95\% c.l.) on the mass of the heavy weak bosons $M_{W^H}$ (as a function of
  mixing parameter $\sin^2\phi$) by electroweak data \cite{NCETC}.
  Below the dashed line, the predicted value of $\Delta R_\sigma/
  R_\sigma \geq 24$\% .}
\label{ncetclim}
\end{figure}

What allows the corrections to single top-quark production to be
relatively large in non-commuting ETC models is the fact that there is
no direct ETC effect on the $Wtb$ vertex to cancel the contributions
from weak gauge boson mixing.  This is in contrast to the calculation of
$R_b$, where such a cancelation does occur.  Hence within the context
of these models it is possible for $R_b$ to have a value close to the
standard model prediction while $R_\sigma$ is visibly altered.

\section{Models with extended weak gauge groups}
\label{sec:weakly}
\setcounter{equation}{0}

There are also models with extended electroweak gauge groups (but no
technicolor sector) that predict an $R_\sigma$ that differs from the
standard model value\footnote{The Left-Right symmetric model \cite{lrm}
  is not among them.  In the limit of no mixing between the left and
  right-handed $W$ bosons, $R_\sigma$ would have the standard model
  value.  The experimentally allowed mixing is small so that including
  mixing should not qualitatively alter the conclusion.  The extra $W$
  boson in the alternative left-right model \cite{alrm} carries lepton
  number and would not contribute to single top production.} The
analysis of weak gauge boson mixing presented in section 3.2 can be
adapted to these models.

\subsection{topflavor}

A recently-introduced model known as topflavor 
\cite{topflavor}\cite{rose} has the same $SU(2)_{heavy} \times
SU(2)_{light} \times U(1)_{Y}$ electroweak gauge group as non-commuting
ETC (without an underlying ETC sector).  Again, the third generation of
fermions couples to $SU(2)_{heavy}$ while the first and second
generations couple to $SU(2)_{light}$.  The simplest forms of the
symmetry-breaking sector include a scalar which transforms as $(2,2)_0$
and one which is a doublet under only one of the $SU(2)$ groups.  As in
non-commuting ETC, there are therefore `heavy' and `light' cases of
topflavor according to whether the second scalar transforms as a doublet
under $SU(2)_{heavy}$ or $SU(2)_{light}$ (i.e.  according to whether the
same order parameter gives mass to the weak gauge bosons and the heavy
fermions).  The phenomenology of the heavy case is explored in ref. 
\cite{topflavor} and that of the light case is discussed in refs. 
\cite{rose} and \cite{topflavor2}.

The analysis of topflavor is similar to that of non-commuting extended
technicolor.  The calculated value of $\Delta R_\sigma/ R_\sigma$ is the
same since the weak sectors of the two models are identical.  It is the
experimental constraints on the models' parameter spaces that differ
(since the non-commuting ETC model contains parameters not present in
topflavor).

We can find a lower bound on the allowed value of the heavy $W$ mass in
heavy-case topflavor by realizing that the extra $W$ boson causes a
fractional shift in $R_{\mu\tau}$, just as in non-commuting ETC 
\cite{NCETC}
\be
(\Delta R_{\mu\tau})^{topflavor}_{heavy} = - 2 / x \iip
\ee
Since current experiment \cite{rmutdat} requires $\vert \Delta
R_{\mu\tau}\vert \le 1.8$\% at the $2\sigma$ level, we can apply
equation (\ref{mwzrat}) to find the lower bound
\be
M_{W^H} \ge 10.5\, M_W / s c
\ee
on the heavy $W$ boson's mass.  When this bound is satisfied, the value
of $\vert\Delta R_\sigma/ R_\sigma\vert$ always lies below\footnote{This
  maximum fractional shift in $R_\sigma$ is obtained when $\sin^2\phi$
  is at its minimum value of 0.034.  A smaller value of $\sin^2\phi$
  would make $g_{light}$ large enough to break the light fermions'
  chiral symmetries.  The critical value of the coupling is estimated
  using the results of a gap-equation analysis of chiral symmetry
  breaking in the ``rainbow'' approximation \cite{rainbow}; see \cite{NCETC} for
  further details.} 13.5\%, so that the change in the rate of single top
quark production is not likely to be visible at the Tevatron.

The current experimental constraints for the light case of topflavor
have been explored in \cite{rose}.  When the constraints are expressed
as a lower bound on the mass of the extra weak bosons (as a function of
mixing parameter $\sin^2\phi$), they appear stronger than those on
non-commuting ETC.  In other words, the shape of the exclusion curve is
similar to that shown in Figure 1, but lies above it, with the lowest
allowed value of $M_{W^H}$ being about 1.1 TeV.  As a result, the change
in the rate of single top quark production in the light case of
topflavor always lies below about 13\%.  Again, this is unlikely to be
observable.

\subsection{Ununified standard model}

The ununified standard model \cite{uum} also sports an extended weak
gauge group with two $SU(2)$ components and a single $U(1)$.  However,
in this case, the quarks transform according to one non-abelian group
($SU(2)_q$) and the leptons according to the other ($SU(2)_{\ell}$).
In order to preserve the experimentally verified relationship between
the leptonic and semi-leptonic weak interactions that holds in the
standard model, the symmetry-breaking sector must be of 
the ``light'' type in which no new low-energy charged current
interactions between a leptonic and a hadronic current occur.  The
simplest possibility is therefore to have one scalar that transforms
as a (2,2) under the two $SU(2)$ groups and another that is an
$SU(2)_{\ell}$ doublet, but an $SU(2)_q$ singlet.

The extra weak gauge boson mixing angle $\phi_{uum}$ in this model is
conventionally defined so that $\sin\phi_{uum} \leftrightarrow
\cos\phi_{NC-ETC}$.  Otherwise, the formalism developed earlier for
the analysis of non-commuting ETC carries through; explicit expressions
for the top-bottom and leptonic cross-sections are in the Appendix.

A fit of the ununified standard model to precision electroweak data 
\cite{uumlim} has found a 95\% c.l. lower bound of just under 2 TeV on the
masses of the heavy $W$ and $Z$ bosons\footnote{This is the bound for
  zero mixing angle; the bound gets even stronger as $\sin^2\phi_{uum}$
  increases.}.  Keeping this in mind, and restricting the value of
$\sin^2\phi_{uum}$ to exceed the critical value of 0.034, we
checked for an intersection between the experimentally allowed parameter
space and the region of visible alteration of the $Wtb$ vertex.

We find a small region in the $\sin^2\phi_{uum} - M_{W^H}$ plane, the
shaded triangle in Figure 2, which is allowed by experiment and in which
$\Delta R_\sigma/ R_\sigma \leq - 16\%$.  Elsewhere in the model's
experimentally parameter space, the shift in $R_\sigma$ is too small to
be reliably detected by an experimental precision of $\pm 8$\% .  Note
that since the shift is negative, it is distinct from that predicted by
models like non-commuting ETC which have an $SU(2)_{heavy} \times
SU(2)_{light}$ group structure.  

Furthermore, $R_b$ has essentially the standard
model value in the region where $\Delta R_\sigma/ R_\sigma$ is large.  
One may calculate the shift in $R_b$ by repeating the analysis of
section 3.2 for the $Z$ bosons and finding how $Zq\bar q$ couplings are
altered.  The result \cite{uumlim} is that $\Delta R_b / R_b \approx -.052
(M_W/M_{W^H})^2 / c^2$.  Since $c^2 \geq .83$ and $M_{W^H} \gae 2$ TeV
in the region in question, $\vert \Delta R_b / R_b \vert \lae 10^{-4}$.
Qualitatively this is because no factor of the top quark mass enters to
enhance the shift in $R_b$ as can happen in ETC models.

\begin{figure}[htb]
\epsfxsize 10cm \centerline{\epsffile{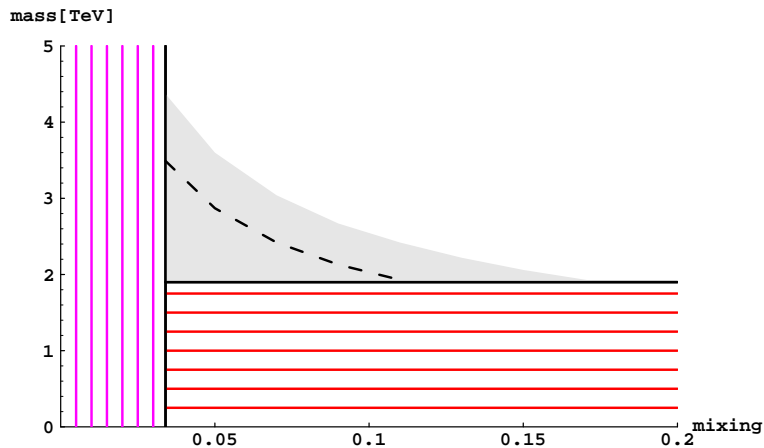}}
\caption[uumlim]{Region (shaded) where the ununified standard model
  predicts a visible decrease($\Delta R_\sigma/ R_\sigma \leq
  - 16$\%) in single top quark production at TeV33. Below the dashed
  line, the predicted decrease is $\Delta R_\sigma/ R_\sigma \leq
  - 24$\%.  The horizontally
  hatched region marks the lower bound on the mass of the heavy weak bosons $M_{W^H}$
  (for small mixing parameter $\sin^2\phi$) from electroweak
  data \cite{NCETC}. In the vertically hatched region, the chiral
  symmetries of the fermions would be broken by a strong weak coupling.} 
\label{uumlim}
\end{figure}

\section{Discussion}
\label{sec:concl}
\setcounter{equation}{0}

Measuring the rate of single top-quark production in Run 3 at the
Tevatron offers a promising opportunity to test models of electroweak physics.  
We have shown here that models with extra $W$ bosons can predict an
alteration of $R_\sigma$ that would be visible to experiment, provided
that the new $W$ bosons weigh less than a few  TeV.

In particular we found interesting results for models with an
$SU(2)_{heavy}\times SU(2)_{light}$ weak gauge group and an electroweak
symmetry breaking condensate charged under the {\bf light} rather
than the {\bf heavy} $SU(2)$.  In such models, the value of
$R_\sigma$ can be greatly increased above the standard model prediction.
Hence the value of $R_\sigma$ provides a valuable test of the dynamical
symmetry-breaking models involving non-commuting extended technicolor.  If
the measurement attains a greater precision than assumed here,
it may also be possible to test the related model with fundamental
scalars known as topflavor.

The predicted increase in $R_\sigma$ is not only visible, but
distinctive.  As we have seen, other models with extra weak bosons that
can alter $R_\sigma$ predict either a shift that is too small to be seen
(e.g. ordinary ETC, topcolor-assisted technicolor, left-right-symmetric
model, heavy-case non-commuting ETC or topflavor) or a shift towards a
lower value of $R_\sigma$ (e.g.  the ununified standard model). This
trend continues when models including other kinds of non-standard
physics are examined.  Adding a fourth generation of quarks would tend
to reduce $\vert V_{tb}\vert$ and, thus, $R_\sigma$.  The extra scalar
bosons in 2-Higgs-Doublet models \cite{2hdref} have been found
\cite{lioky} to reduce $R_\sigma$ by an amount not greater than 15\% .
The electroweak contributions in the minimal supersymmetric standard
model \cite{mssmref} likewise reduce alter $R_\sigma$ by no more than
$\pm$10\% \cite{lioky2} (the sign varies over the model's parameter
space).

\bigskip
\centerline{\bf Acknowledgments}
\vspace{12pt}

We thank R.S. Chivukula and K.D. Lane for useful conversations and
comments on the manuscript.  E.H.S. also acknowledges the support of the
NSF Faculty Early Career Development (CAREER) program and the DOE
Outstanding Junior Investigator program. {\em This work was supported in
  part by the National Science Foundation under grant PHY-95-1249 and by
  the Department of Energy under grant DE-FG02-91ER40676.}

\bigskip
\centerline{\bf Appendix}

Here we present some details of our calculation for the reader's
convenience.  The cross-section for production of a fermion/anti-fermion
pair via exchange of $W^L$ and $W^H$ bosons contains the following terms
\bea
\left[C_f \hat{u}(\hat{u} - m_f^2)\right]&&\!\!\!\!\! \times\ 
 \left[ {\alpha\over (\hat{s} - M_W^2)}\ \  + \right. \\
 &&\left. {2\ \beta\ (\hat{s} - M_{W^H}^2) \over {\hat{s}((\hat{s} -
      M_{W^H}^2)^2 + \Gamma_{W^H}^2 M_{W^H}^2)}} + {\gamma\over{((\hat{s} -
      M_{W^H}^2)^2 + \Gamma_{W^H}^2 M_{W^H}^2)}} \right] \nonumber
\eea
where $m_f$ is $m_t$ for the $tb$ final state and zero for the $l\nu$
final state, $C_f$ is 3 for the $tb$ final state and 1 for the $l\nu$
final state, $V_{tb}$ has been set equal to 1, and multiplicative
constants which cancel in the ratio $R_\sigma$ have been dropped.  Here
$\Gamma_{W^H}$ is taken to be the s-dependent width of the heavy weak
boson so that the results match correctly onto those from calculations
based on four-fermion operators.

The coefficients $\alpha$, $\beta$, and $\gamma$ are specific to the
process ($tb$ or $l\nu$ production) and the model.  We write them in
terms of the heavy $W$ boson mass $M_{W^H}$ and the weak boson mixing
angle ($s \equiv \sin\phi,\ c \equiv \cos\phi$).  They have been derived
using equations (\ref{stnst}) (\ref{dgzi}) and (\ref{dgzii}) and
dropping terms of order $x^{-2}$ or higher (where $x \equiv u^2/v^2$ is
the ratio of mixing and breaking vevs-squared).
In the heavy case of non-commuting ETC or topflavor:
\bea
\alpha^{tb} &=& -\beta^{tb} = \gamma^{tb} = 1 + {2(c^2-s^2)\over{c^2}} \left(
  M_W^2 \over M_{W^H}^2 \right) \equiv \alpha^{tb}_h \nonumber\\ 
\alpha^{l\nu} &=& 1 + 4{M_W^2\over M_{W^H}^2} \equiv \alpha^{l\nu}_h
  \nonumber \\
\beta^{l\nu} &=&  {c^2\over s^2} +
  {2(c^2-s^2)\over{c^2}} \left(M_W^2\over M_{W^H}^2 \right)  \equiv
  \beta^{l\nu}_h \nonumber \\  
\gamma^{l\nu}  &=& {c^4\over s^4} - {4 c^2\over
  s^2}  \left( M_W^2\over M_{W^H}^2 \right) \equiv \gamma^{l\nu}_h \iip
\eea
For the light case of non-commuting ETC or topflavor
\bea
\alpha^{tb} &=& -\beta^{tb} = \gamma^{tb} = 1 + {2(s^2-c^2)\over{s^2}} \left( {M_W^2\over
M_{W^H}^2} \right) \equiv \alpha^{tb}_l \nonumber\\
\alpha^{l\nu}  &=& 1 - 4{c^2\over s^2}{M_W^2\over M_{W^H}^2}
\nonumber  \\
\beta^{l\nu}  &=&  {c^2\over s^2} - {2 c^2\over{s^4}} \left( 
 { M_W^2\over M_{W^H}^2} \right)\nonumber \\
\gamma^{l\nu} &=& {c^4\over s^4} - {4 c^4\over s^4} \left({ M_W^2\over
M_{W^H}^2} \right)  \iip
\eea
In the ununified standard model:
\bea
\alpha^{tb} &=& \alpha^{l\nu}_h\ \ \ \ \ \ \ \ \ \ \ \ 
\alpha^{l\nu} = \alpha^{tb}_h  \nonumber\\
\beta^{tb} &=& \beta^{l\nu}_h\ \ \ \ \ \ \ \ \ \ \ \ \beta^{l\nu}
= -\alpha^{tb}_h \nonumber\\
\gamma^{tb} &=& \gamma^{l\nu}_h \ \ \ \ \ \ \ \ \ \ \ \ 
\gamma^{l\nu}  = \alpha^{tb}_h \iip
\eea

To find the hadronic cross-section for each process, we used MRSDO'
structure functions and integrated over
center-of-mass energy ( $m_t + m_b < \sqrt{\hat{s}} < 1 {\rm TeV}$) boost
rapidity ($-2.0 < Y_{boost} < 2.0$), and center-of-mass scattering angle
(to the kinematic limit imposed by the masses and greatest rapidity
($\pm 2.0$) of the final state particles).  Our results were insensitive
to the precise choice of energy and rapidity integration limits.


\end{document}